\documentclass[aps,prb,preprint,groupedaddress,amsmath]{revtex4}
\usepackage{graphicx}

\begin{document}

\title{Finite element simulation of microphotonic lasing system}

\author{Chris Fietz$^{1}$ and Costas Soukoulis$^{1,2}$}
\affiliation{$^1$Ames Laboratory and Department of Physics and Astronomy, Iowa State University, Ames, Iowa 50011, USA \\ $^2$Institute of Electronic Structure and Laser, FORTH, 71110 Heraklion, Crete, Greece}

\begin{abstract}
We present a method for performing time domain simulations of a microphotonic system containing a four level gain medium based on the finite element method.  This method includes an approximation that involves expanding the pump and probe electromagnetic fields around their respective carrier frequencies, providing a dramatic speedup of the time evolution.  Finally, we present a two dimensional example of this model, simulating a cylindrical spaser array consisting of a four level gain medium inside of a metal shell.
\end{abstract}

\maketitle

\section{Introduction}
Interest in microphotonic lasing systems has been increasing over the past few years.  As a result, it has become more important to be able to numerically simulate these lasing systems.  Several finite difference time domain (FDTD) simulations of a four level gain medium embedded in a microphotonic system have been presented previously~\cite{Bohringer_08,Fang_09,Fang_PHD,Fang_10a,Fang_10b,Wuestner_10,Fang_11,Wuestner_11,Hamm_11}, but these simulations all use structured (cubic) grids and consequently accurately model curved geometries.  There have also been methods developed to model spherical gain geometries by expanding electromagnetic fields as sums of spherical Bessel functions~\cite{Gordon_07,Gordon_08}.  These methods overcome the limitations of structured grids for spherical geometries, but in turn are limited to only modelling spherical geometries.  In principle it should be possible to model a microphotonic lasing system with an FDTD simulation utilizing unstructured grids, but to the best knowledge of the authors this has not been demonstrated.  The finite element method (FEM) can utilize unstructured grids and as a result can model a wide variety of geometries.  In this paper we present a FEM model of a microphotonic system with gain arising from a four level quantum system.  In addition to developing a FEM microphotonic lasing model, an approximation is introduced whereby the pump and probe fields are solved for separately, with each field described by the slowly varying complex valued field amplitude of a constant frequency carrier wave.  This approximation allows for much larger time steps and a considerable speedup in simulation time.

In the first section of this paper we will describe the dynamics of the microphotonic lasing system, the carrier wave approximation, and finally the finite element formulation of the problem.  In the second section we present a two dimensional model of a one dimensional cylindrical spaser array as an example of this new simulation method.

\section{FEM microphotonic lasing simulation}
\subsection{Field equations of a microphotonic lasing system}

The simulation we present of a microphotonic lasing system requires the time domain modelling of several different fields and their mutual interactions.  These fields include the electromagnetic field, the electric polarization field inside a metal with a Drude response, the electric polarization field of the gain medium, and the population density fields of the different energy levels of the gain medium.  Each of these fields evolve according a particular differential equation that must be solved when simulating a microphotonic lasing system. The field equation for the electromagnetic field is

\begin{equation}\label{F_field1}
\nabla\times\left( \displaystyle\frac{1}{\mu_0}\nabla\times\textbf{A} \right) + \epsilon_r\epsilon_0\frac{\partial^2 \textbf{A}}{\partial t^2} = \frac{\partial \textbf{P}}{\partial t},
\end{equation}

\noindent where $\textbf{A}$  is the electromagnetic vector potential, $\textbf{P}$ is a polarization vector describing either a
Drude response from a metal inclusion or a Lorentzian response from a four level gain system, and $\epsilon_r$ is a relative permittivity that is constant with respect to frequency and not included in $\textbf{P}$.  Here and for the remainder of the paper we have use SI units.  Also, we have used the temporal gauge condition $\partial \mathrm{A}_0/\partial t=0$ along with the initial condition for the
electrostatic potential $\mathrm{A}_0(t=0,\textbf{x}) = 0$ to ensure that the electrostatic potential is zero for all time, eliminating it from
our equations. Given this choice of gauge the electric field and magnetic flux density are defined as

\begin{equation}
\begin{array}{c}
\textbf{E} = -\displaystyle\frac{\partial\textbf{A}}{\partial t}, \\ \\
\textbf{B} = \nabla\times\textbf{A}
\end{array}
\end{equation}

\noindent Using this definition for $\textbf{E}$, the Drude response of a metal inclusion is determined by the equation

\begin{equation}\label{F_field2}
\displaystyle\frac{\partial \textbf{P}}{\partial t} + \gamma\textbf{P} = - \epsilon_0\omega_p^2\textbf{A}
\end{equation}

\noindent where $\omega_p$ is the plasma frequency and $\gamma$ is the damping frequency of the Drude metal.

\begin{figure}[!h]
\begin{center}
\includegraphics[width=0.5\textwidth]{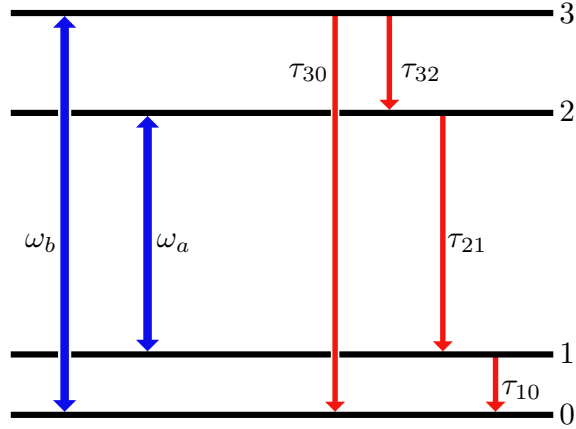}
\end{center}
\caption{Simple model of a four level gain medium.  The lasing and pump transitions are assumed to be electric dipole transistions with frequencies of $\omega_a$ and $\omega_b$ respectively.  The decay processes between the i-th and j-th energy levels are described by the decay rates $1/\tau_{ij}$.}\label{Fig_1}
\end{figure}

The gain medium is modelled as simple four level quantum system, described schematically in Fig.~\ref{Fig_1}.  The $1 \rightarrow 2$ transition is an electric dipole transition with a frequency of $\omega_a$.  Similarly, the $0 \rightarrow 3$ transition is also an electric dipole transition with frequency $\omega_b$.  Spontaneous decay between the i-th level to the j-th level occurs at the decay rate of $1/\tau_{ij}$.  These decay rates include both radiative (spontaneous photon emission) and non-radiative (spontaneous phonon emission) decay processes.  In the case of spontaneous photon emission, our model does not produce a photon.  Coupling of the gain medium to the electromagnetic field is only allowed for stimulated photon emission.

The electromagnetic response of the four level gain system is given by

\begin{equation}\label{F_field3}
\begin{array}{rcl}
\displaystyle\frac{\partial^2 \mathrm{P}_{ai}}{\partial t^2} + \Gamma_a\frac{\partial \mathrm{P}_{ai}}{\partial t} + \omega_a^2 \mathrm{P}_{ai} & = & -\sigma_a(\mathrm{N}_{2i} - \mathrm{N}_{1i})\mathrm{E}_i, \\ \\
\displaystyle\frac{\partial^2 \mathrm{P}_{bi}}{\partial t^2} + \Gamma_b\frac{\partial \mathrm{P}_{bi}}{\partial t} + \omega_b^2 \mathrm{P}_{bi} & = & -\sigma_b(\mathrm{N}_{3i} - \mathrm{N}_{0i})\mathrm{E}_i.
\end{array}
\end{equation}

\noindent Here $\mathrm{P}_i^a$ and $\mathrm{P}_i^b$ are the i-th components of the gain polarization due to transitions between the 1st
and 2nd levels and between the 0th and 3rd levels respectively.  Additionally, $\Gamma_a$ and $\Gamma_b$ are the linewidths of
these transitions, $\sigma_a$ and $\sigma_b$ are coupling constants, and $\mathrm{N}_{0i}$, $\mathrm{N}_{1i}$, $\mathrm{N}_{2i}$ and $\mathrm{N}_{3i}$ are the population number densities for oscillators polarized in the i-th direction for the 0th, 1st, 2nd and 3rd
energy levels.  Note that $\Gamma_a\geq1/\tau_{21}$ and $\Gamma_b\geq1/\tau_{30}$~\cite{Siegman_86}.

Finally, the population number densities evolve according the equations \cite{Siegman_86,Jiang_00}

\begin{equation}\label{F_field4}
\begin{array}{rcl}
\displaystyle\frac{\partial \mathrm{N}_{3i}}{\partial t} & = & \displaystyle\frac{1}{\hbar\omega_b} \mathrm{E}_i\frac{\partial \mathrm{P}_{bi}}{\partial t} - \left(\frac{1}{\tau_{30}}+\frac{1}{\tau_{32}}\right)\mathrm{N}_{3i}, \\ \\
\displaystyle\frac{\partial \mathrm{N}_{2i}}{\partial t} & = & \displaystyle\frac{\mathrm{N}_{3i}}{\tau_{32}} + \frac{1}{\hbar\omega_a} \mathrm{E}_i\frac{\partial \mathrm{P}_{ai}}{\partial t} - \frac{\mathrm{N}_{2i}}{\tau_{21}}, \\ \\
\displaystyle\frac{\partial \mathrm{N}_{1i}}{\partial t} & = & \displaystyle\frac{\mathrm{N}_{2i}}{\tau_{21}} - \frac{1}{\hbar\omega_a} \mathrm{E}_i\frac{\partial \mathrm{P}_{ai}}{\partial t} - \frac{\mathrm{N}_{1i}}{\tau_{10}}, \\ \\
\displaystyle\frac{\partial \mathrm{N}_{0i}}{\partial t} & = & \displaystyle\frac{\mathrm{N}_{3i}}{\tau_{30}} + \displaystyle\frac{\mathrm{N}_{1i}}{\tau_{10}} - \frac{1}{\hbar\omega_b} \mathrm{E}_i\frac{\partial \mathrm{P}_{bi}}{\partial t},
\end{array}
\end{equation}

Together, this system of equations (Eqs.~(\ref{F_field1},\ref{F_field2}-\ref{F_field4})) completely describes the dynamics of the microphotonic lasing system. The main disadvantage of solving this system of differential equations is the small time step required. In practice, $\sim 100$ time steps per period of the pumping laser beam are required for an adequate simulation. A typical lasing simulation could require over 100,000 lasing periods, making the computational requirements of the simulation prohibitively large.

\subsection{Period averaged approximation}

There is a simple method for dramatically speeding up the simulation time. The electromagnetic field as well as the polarization fields oscillates at two frequencies. These two frequencies are approximately equal to the frequency of the $1 \rightarrow 2$ transition frequency $\omega_a$, and the $0 \rightarrow 3$ transition frequency $\omega_b$.  Much of the computational effort required in this time domain simulation is spent on these simple, approximately harmonic oscillations. A good approximation is to assume these fields oscillate harmonically, with complex valued amplitudes that are slowly changing in time. We can ignore the fast oscillations and instead simulate the relatively slower time dependence of these amplitudes.

Since there are two frequencies, we divide our electromagnetic field into two separate fields

\begin{equation}
\textbf{A}(t,\textbf{x}) = \displaystyle\frac{\textbf{A}_1(t,\textbf{x})e^{\mathrm{i}\omega_1 t} + \textbf{A}_2(t,\textbf{x})e^{\mathrm{i}\omega_2 t} + c.c.}{2},
\end{equation}

\noindent with each field oscillating at a different frequency. Here $\textbf{A}_1$ is the complex valued amplitude for an electromagnetic field that oscillates at a frequency close to the $1 \rightarrow 2$ transition ($\omega_1 \approx \omega_a$), and $\textbf{A}_2$ is the complex valued amplitude for an electromagnetic field that oscillates close to the $0 \rightarrow 3$ transition ($\omega_2 \approx \omega_b$).  Also, $c.c.$ indicates the complex conjugate of the preceding terms.

By inserting the above equation into Eq.~(\ref{F_field1}), the field equation for $\textbf{A}$, we derive two new field equations

\begin{equation}\label{T_field1}
\begin{array}{c}
\nabla\times\left( \displaystyle\frac{1}{\mu_0}\nabla\times\textbf{A}_1 \right) + \epsilon_r\epsilon_0\left(\displaystyle  -\omega_1^2\textbf{A}_1 + 2\mathrm{i}\omega_1\frac{\partial\textbf{A}_1}{\partial t} + \frac{\partial^2 \textbf{A}_1}{\partial t^2} \right) = \displaystyle\frac{\partial\textbf{P}_1}{\partial t}, \\ \\
\nabla\times\left( \displaystyle\frac{1}{\mu_0}\nabla\times\textbf{A}_2 \right) + \epsilon_r\epsilon_0\left( \displaystyle -\omega_2^2\textbf{A}_2 + 2\mathrm{i}\omega_2\frac{\partial\textbf{A}_2}{\partial t} + \frac{\partial^2 \textbf{A}_2}{\partial t^2} \right) = \displaystyle\frac{\partial\textbf{P}_2}{\partial t}.
\end{array}
\end{equation}

\noindent Here we have also separated the polarization field into two fields

\begin{equation}
\textbf{P}(t,\textbf{x}) = \displaystyle\frac{\textbf{P}_1(t,\textbf{x})e^{\mathrm{i}\omega_1 t} + \textbf{P}_2(t,\textbf{x}) e^{\mathrm{i}\omega_2 t} + c.c}{2}.
\end{equation}

\noindent For Drude metal inclusions the polarization fields obey the equations

\begin{equation}\label{T_field2}
\begin{array}{c}
\mathrm{i}\omega_1\textbf{P}_1^d + \displaystyle\frac{\partial\textbf{P}_1^d}{\partial t} + \gamma\textbf{P}_1^d = -\epsilon_0\omega_p^2\textbf{A}_1, \\ \\
\mathrm{i}\omega_2\textbf{P}_2^d + \displaystyle\frac{\partial\textbf{P}_2^d}{\partial t} + \gamma\textbf{P}_2^d = -\epsilon_0\omega_p^2\textbf{A}_2,
\end{array}
\end{equation}

\noindent while for Lorentzian gain inclusions the polarization fields obey the equations

\begin{equation}\label{T_field3}
\begin{array}{c}
-\omega_1^2\mathrm{P}_{1i}^g + 2\mathrm{i}\omega_1\displaystyle\frac{\partial\mathrm{P}_{1i}^g}{\partial t} + \frac{\partial^2\mathrm{P}_{1i}^g}{\partial t^2} + \Gamma_a\left( \mathrm{i}\omega_1\mathrm{P}_{1i}^g + \frac{\partial\mathrm{P}_{1i}^g}{\partial t} \right) + \omega_a^2\mathrm{P}_{1i}^g = -\sigma_a\left(\mathrm{N}_{2i} - \mathrm{N}_{1i}\right)\mathrm{E}_{1i}, \\ \\
-\omega_2^2\mathrm{P}_{2i}^g + 2\mathrm{i}\omega_2\displaystyle\frac{\partial\mathrm{P}_{2i}^g}{\partial t} + \frac{\partial^2\mathrm{P}_{2i}^g}{\partial t^2} + \Gamma_b\left( \mathrm{i}\omega_2\mathrm{P}_{2i}^g + \frac{\partial\mathrm{P}_{2i}^g}{\partial t} \right) + \omega_b^2\mathrm{P}_{2i}^g = -\sigma_b\left(\mathrm{N}_{3i} - \mathrm{N}_{0i}\right)\mathrm{E}_{2i}.
\end{array}
\end{equation}

\noindent Here $\mathrm{E}_{1i}$ and $\mathrm{E}_{2i}$ are the i-th components of the electric fields associated with the potentials $\textbf{A}_1$ and $\textbf{A}_2$, and are defined as $\textbf{E}_1=-\partial\textbf{A}_1/\partial t$ and $\textbf{E}_2 = -\partial\textbf{A}_1/\partial t$ respectively.

Finally, the new differential equations for the occupation number densities are

\begin{equation}\label{T_field4}
\begin{array}{cl}
\displaystyle\frac{\partial \mathrm{N}_{3i}}{\partial t} =& \displaystyle\frac{1}{\hbar\omega_b} \left\langle \mathrm{E}_{2i}\frac{\partial\mathrm{P}_{2i}}{\partial t} \right\rangle - \left(\frac{1}{\tau_{30}}+\frac{1}{\tau_{32}}\right)\mathrm{N}_{3i} , \\ \\
\displaystyle\frac{\partial \mathrm{N}_{2i}}{\partial t} =& \displaystyle\frac{\mathrm{N}_{3i}}{\tau_{32}} + \frac{1}{\hbar\omega_a} \left\langle \mathrm{E}_{1i}\frac{\partial\mathrm{P}_{1i}}{\partial t} \right\rangle - \frac{\mathrm{N}_{2i}}{\tau_{21}} , \\ \\
\displaystyle\frac{\partial \mathrm{N}_{1i}}{\partial t} =& \displaystyle\frac{\mathrm{N}_{2i}}{\tau_{21}} - \frac{1}{\hbar\omega_a} \left\langle \mathrm{E}_{1i}\frac{\partial\mathrm{P}_{1i}}{\partial t} \right\rangle - \frac{\mathrm{N}_{1i}}{\tau_{10}} , \\ \\
\displaystyle\frac{\partial \mathrm{N}_{0i}}{\partial t} =& \displaystyle\frac{\mathrm{N}_{3i}}{\tau_{30}} + \frac{\mathrm{N}_{1i}}{\tau_{10}} - \frac{1}{\hbar\omega_b} \left\langle \mathrm{E}_{2i}\frac{\partial\mathrm{P}_{2i}}{\partial t} \right\rangle .
\end{array}
\end{equation}

\noindent Here the coupling term between the occupation number density fields and the electromagnetic and
polarization fields has been replaced by a term representing the period averaged value of these terms

\begin{equation}\label{T_field5}
\begin{array}{c}
\left\langle\mathrm{E}_{1i}\displaystyle\frac{\partial\mathrm{P}_{1i}}{\partial t}\right\rangle = \displaystyle\frac{1}{2}\mathrm{Re}\left[\left(-\mathrm{i}\omega_1\mathrm{A}_{1i}-\frac{\partial\mathrm{A}_{1i}}{\partial t}\right)^*\left(\mathrm{i}\omega_1\mathrm{P}_{1i} + \frac{\partial\mathrm{P}_{1i}}{\partial t}\right)\right], \\ \\
\left\langle\mathrm{E}_{2i}\displaystyle\frac{\partial\mathrm{P}_{2i}}{\partial t}\right\rangle = \displaystyle\frac{1}{2}\mathrm{Re}\left[\left(-\mathrm{i}\omega_1\mathrm{A}_{2i}-\frac{\partial\mathrm{A}_{2i}}{\partial t}\right)^*\left(\mathrm{i}\omega_1\mathrm{P}_{2i} + \frac{\partial\mathrm{P}_{2i}}{\partial t}\right)\right],
\end{array}
\end{equation}

\noindent where $*$ indicates complex conjugation.

Finally, we mention that while we have developed the preceding approximation using a FEM model, this approximation is not limited to the FEM.  It could potentially be used to speedup both the FDTD models of microphotonic lasing systems~\cite{Bohringer_08,Fang_09,Fang_PHD,Fang_10a,Fang_10b,Wuestner_10,Fang_11,Wuestner_11,Hamm_11} as well as the time domain models of spherical lasing geometries utilizing spherical Bessel functions~\cite{Gordon_07,Gordon_08}.

\subsection{Finite element formulation}\label{Sec_2_3}

Now that we have derived the period averaged field equations (Eqs.~(\ref{T_field1},\ref{T_field2}-\ref{T_field5})) for the microphotonic lasing system, we can convert these differential equations into weak forms that can be solved in a finite element simulation.

The weak forms for the field equations of the electromagnetic fields (Eq.~(\ref{T_field1})) are

\begin{equation}\label{W_field1}
\begin{array}{rl}
\mathrm{F}_{A1}(\tilde{\textbf{A}}_1,\textbf{A}_1) = &  \left(\nabla\times\tilde{\textbf{A}}_1\right)\cdot \displaystyle\frac{1}{\mu_0} \left(\nabla\times\textbf{A}_1 \right) + \epsilon_r\epsilon_0 \tilde{\textbf{A}}_1\cdot \left(\displaystyle  -\omega_1^2\textbf{A}_1 + 2\mathrm{i}\omega_1\frac{\partial\textbf{A}_1}{\partial t} + \frac{\partial^2 \textbf{A}_1}{\partial t^2} \right) \\ \\
& - \tilde{\textbf{A}}_1\cdot \displaystyle\frac{\partial\textbf{P}_1}{\partial t}, \\ \\
\mathrm{F}_{A2}(\tilde{\textbf{A}}_2,\textbf{A}_2) = & \left(\nabla\times\tilde{\textbf{A}}_2\right)\cdot \displaystyle\frac{1}{\mu_0} \left(\nabla\times\textbf{A}_2 \right) + \epsilon_r\epsilon_0 \tilde{\textbf{A}}_2\cdot \left( \displaystyle -\omega_2^2\textbf{A}_2 + 2\mathrm{i}\omega_2\frac{\partial\textbf{A}_2}{\partial t} + \frac{\partial^2 \textbf{A}_2}{\partial t^2} \right) \\ \\
& - \tilde{\textbf{A}}_2\cdot \displaystyle\frac{\partial\textbf{P}_2}{\partial t}.
\end{array}
\end{equation}

\noindent  Here $\sim$ indicates a test function~\cite{Zimmerman_04,Jin_02}.  These weak forms enforce both the electromagnetic field equations as well as a natural boundary condition~\cite{Zimmerman_04,Jin_02}.  The finite element method requires that the integral of the weak form over the simulation domain be set to zero.  As an example, if we apply this requirement to the weak form $\mathrm{F}_{A1}$, we find that by integrating by parts we obtain a volume integral enforcing the electromagnetic field equation as well as a second boundary integral enforcing a boundary condition on the field,

\begin{equation}\label{B1}
\begin{array}{rcl}
0 &=& \displaystyle\int_{\Omega} d^3x \ \ \mathrm{F}_{A1} \\ \\
&=& \displaystyle\int_{\Omega} d^3x \ \ \tilde{\textbf{A}}_1\cdot \left[ \nabla\times\left( \displaystyle\frac{1}{\mu_0}\nabla\times\textbf{A}_1 \right) + \epsilon_r\epsilon_0\left(\displaystyle  -\omega_1^2\textbf{A}_1 + 2\mathrm{i}\omega_1\frac{\partial\textbf{A}_1}{\partial t} + \frac{\partial^2 \textbf{A}_1}{\partial t^2} \right) - \displaystyle\frac{\partial\textbf{P}_1}{\partial t} \right] \\ \\
&&- \displaystyle\oint_{\partial\Omega} dA \ \ \tilde{\textbf{A}}_1\cdot \left[\hat{\textbf{n}}\times\left(\displaystyle\frac{1}{\mu_0}\nabla\times\textbf{A}_1\right)\right].
\end{array}
\end{equation}

\noindent  Here $\Omega$ is the simulation domain, $\partial \Omega$ is the boundary of that domain, $dA$ is a infinitesimal differential area on that boundary, and $\hat{\textbf{n}}$ is the direction normal to the boundary.  In the absence of any extra boundary terms, the boundary integral in Eq.~(\ref{B1}) forces the tangential component of the magnetic field $\textbf{H}_1$ to zero.  This \emph{perfect magnetic conductor} boundary condition is not desirable for our simulation, so we will modify it to allow for a boundary that absorbs and emits plane waves at normal incidence to the boundary.

For a flat boundary at a large enough distance from the inclusions in the simulation domain that evanescent waves are negligibly small, if the remaining  propagating fields are normal to this flat boundary then the vector potential can be represented as the sum of two vector potentials,

\begin{equation}\label{plane_waves}
\textbf{A}_1(t,\textbf{x}) = \textbf{a}\left(t-\displaystyle\frac{\hat{\textbf{n}}\cdot\textbf{x}}{c}\right)+\textbf{b}\left(t+\frac{\hat{\textbf{n}}\cdot\textbf{x}}{c}\right).
\end{equation}

\noindent Here $\textbf{a}$ is the vector potential of a plane wave propagating toward the boundary and $\textbf{b}$ is the vector potential of a plane wave propagating away from the boundary.  The boundary condition we desire is one that absorbs $\textbf{a}$ and emits an arbitrarily defined $\textbf{b}$.  If we take the part of the surface integrand from Eq.~\ref{B1} that is within the brackets and substitute Eq.~\ref{plane_waves} for $\textbf{A}_1$ we get

\begin{equation}\label{Boundary_eq}
\begin{array}{cl}
\hat{\textbf{n}}\times\left(\displaystyle\frac{1}{\mu_0}\nabla\times\textbf{A}_1\right) &=\hat{\textbf{n}}\times\biggl( -\displaystyle\frac{\hat{\textbf{n}}}{\mu_0 c}\times\frac{\partial \textbf{a}}{\partial t} + \frac{\hat{\textbf{n}}}{\mu_0 c}\times\frac{\partial \textbf{b}}{\partial t} \biggr) \\ \\
&= \displaystyle\frac{1}{z_0}\biggl( \hat{\textbf{n}}\times\hat{\textbf{n}}\times\left( \textbf{E}_1^{out} - \textbf{E}_1^{inc} \right)\biggl) \\ \\
&= -\displaystyle\frac{1}{z_0} \hat{\textbf{n}}\times\hat{\textbf{n}}\times\biggl( \frac{\partial \textbf{A}_1}{\partial t} + 2\textbf{E}_1^{inc}\biggl),
\end{array}
\end{equation}

\noindent  where $\textbf{E}_1^{out}=-\partial\textbf{a}/\partial t$ is the part of the electric field associated the plane wave $\textbf{a}$ propagating toward the boundary, and $\textbf{E}_1^{inc}=-\partial\textbf{b}/\partial t$ is the part of the electric field associated with the plane wave $\textbf{b}$ propagating away from the boundary, the sum of which is $\textbf{E}_1^{out}+\textbf{E}_1^{inc} = \textbf{E}_1 = -\partial\textbf{A}_1/\partial t$.  Also, $z_0\equiv\sqrt{\mu_0/\epsilon_0}$ is the impedance of free space.  Multiplying Eq.~\ref{Boundary_eq} by a test function $\tilde{\textbf{A}}_1$ and integrating over the domain boundary gives us a new boundary weak term

\begin{equation}
B_{A1}(\tilde{\textbf{A}}_1,\textbf{A}_1) = -\displaystyle\oint_{\partial\Omega} dA \ \ \displaystyle\frac{1}{z_0} \tilde{\textbf{A}}_1\cdot \biggl[ \hat{\textbf{n}}\times\hat{\textbf{n}}\times \biggl( \frac{\partial \textbf{A}_1}{\partial t} + 2\textbf{E}_1^{inc} \biggr) \biggr].
\end{equation}

\noindent Adding this additional boundary weak term to specific boundaries enforces a \emph{matched boundary condition} (referred to as an \emph{absorbing boundary condition} in Ref.~\cite{Jin_02}) which allows for plane waves normal to the boundary to be absorbed and for the incident plane wave $\textbf{E}_1^{inc}$ to be emitted into the domain normal to the boundary.  A matched boundary condition for $\textbf{A}_2$ can be enforced in the same manner.

The weak forms for the remaining field equations are simpler since these differential equations only involve derivatives with respect to time.  The weak form for the polarization of Drude metal inclusions is 

\begin{equation}\label{W_field2}
\begin{array}{c}
\mathrm{F}_{PD1}(\tilde{\textbf{P}}_1^d,\textbf{P}_1^d) =  \tilde{\textbf{P}}_1^d \cdot \left[ \mathrm{i}\omega_1\textbf{P}_1^d + \displaystyle\frac{\partial\textbf{P}_1^d}{\partial t} + \gamma\textbf{P}_1^d + \epsilon_0\omega_p^2\textbf{A}_1 \right], \\ \\
\mathrm{F}_{PD2}(\tilde{\textbf{P}}_2^d,\textbf{P}_2^d) =  \tilde{\textbf{P}}_2^d \cdot \left[ \mathrm{i}\omega_2\textbf{P}_2^d + \displaystyle\frac{\partial\textbf{P}_2^d}{\partial t} + \gamma\textbf{P}_2^d + \epsilon_0\omega_p^2\textbf{A}_2 \right],
\end{array}
\end{equation}

\noindent where again a $\sim$ indicates a test function.  Similarly, the weak form for the polarization fields of the gain medium are

\begin{equation}\label{W_field3}
\begin{array}{cl}
\mathrm{F}_{PG1}(\tilde{\textbf{P}}_1^g,\textbf{P}_1^g) = \tilde{\textbf{P}}_1^g \cdot \biggl[ & -\omega_1^2\mathrm{P}_{1i}^g + 2\mathrm{i}\omega_1\displaystyle\frac{\partial\mathrm{P}_{1i}^g}{\partial t} + \frac{\partial^2\mathrm{P}_{1i}^g}{\partial t^2} + \Gamma_a\left( \mathrm{i}\omega_1\mathrm{P}_{1i}^g + \frac{\partial\mathrm{P}_{1i}^g}{\partial t} \right)\biggr. \\ \\
& \biggl. + \omega_a^2\mathrm{P}_{1i}^g + \sigma_a\left(\mathrm{N}_{2i} - \mathrm{N}_{1i}\right)\mathrm{E}_{1i} \biggr], \\ \\
\mathrm{F}_{PG2}(\tilde{\textbf{P}}_2^g,\textbf{P}_2^g) = \tilde{\textbf{P}}_2^g \cdot \biggl[ & -\omega_2^2\mathrm{P}_{2i}^g + 2\mathrm{i}\omega_2\displaystyle\frac{\partial\mathrm{P}_{2i}^g}{\partial t} + \frac{\partial^2\mathrm{P}_{2i}^g}{\partial t^2} +  \Gamma_b\left( \mathrm{i}\omega_2\mathrm{P}_{2i}^g + \frac{\partial\mathrm{P}_{2i}^g}{\partial t} \right) \biggr. \\ \\
&  \biggl. + \omega_b^2\mathrm{P}_{2i}^g + \sigma_b\left(\mathrm{N}_{3i} - \mathrm{N}_{0i}\right)\mathrm{E}_{2i} \biggr],
\end{array}
\end{equation}

\noindent and the weak forms for the population density rate equations are

\begin{equation}\label{W_field4}
\begin{array}{cl}
\mathrm{F}_{N3i}(\tilde{\mathrm{N}}_{3i},\mathrm{N}_{3i}) =& \tilde{\mathrm{N}}_{3i} \cdot \biggl[ \displaystyle\frac{\partial \mathrm{N}_{3i}}{\partial t} - \displaystyle\frac{1}{\hbar\omega_b} \left\langle \mathrm{E}_{2i}\frac{\partial\mathrm{P}_{2i}}{\partial t} \right\rangle + \left(\frac{1}{\tau_{30}}+\frac{1}{\tau_{32}}\right)\mathrm{N}_{3i} \biggr], \\ \\
\mathrm{F}_{N2i}(\tilde{\mathrm{N}}_{2i},\mathrm{N}_{2i}) =& \tilde{\mathrm{N}}_{2i} \cdot \biggl[ \displaystyle\frac{\partial \mathrm{N}_{2i}}{\partial t} - \displaystyle\frac{\mathrm{N}_{3i}}{\tau_{32}} - \frac{1}{\hbar\omega_a} \left\langle \mathrm{E}_{1i}\frac{\partial\mathrm{P}_{1i}}{\partial t} \right\rangle + \frac{\mathrm{N}_{2i}}{\tau_{21}} \biggr], \\ \\
\mathrm{F}_{N1i}(\tilde{\mathrm{N}}_{1i},\mathrm{N}_{1i}) =& \tilde{\mathrm{N}}_{1i} \cdot \biggl[ \displaystyle\frac{\partial \mathrm{N}_{1i}}{\partial t} - \displaystyle\frac{\mathrm{N}_{2i}}{\tau_{21}} + \frac{1}{\hbar\omega_a} \left\langle \mathrm{E}_{1i}\frac{\partial\mathrm{P}_{1i}}{\partial t} \right\rangle + \frac{\mathrm{N}_{1i}}{\tau_{10}} \biggr],
\end{array}
\end{equation}

\noindent where the period averaged values for the coupling term are given in Eq.~(\ref{T_field5}).  Also, we can avoid solving for $\mathrm{N}_{0i}$ by taking advantage of the fact that $\mathrm{N}_{0i}=\mathrm{N}_{int}-\mathrm{N}_{1i}-\mathrm{N}_{2i}-\mathrm{N}_{3i}$ where $\mathrm{N}_{int}$ is the initial value of $\mathrm{N}_{0i}$ when $\mathrm{N}_{1i}=\mathrm{N}_{2i}=\mathrm{N}_{3i}=0$.

\section{Cylindrical spaser array}

As an example of a microphotonic lasing system simulation we present a two dimensional model of a spaser (surface plasmon amplification by stimulated emission of radiation~\cite{Bergman_03,Stockman_08}).  The time domain FEM simulation was performed using the commercial software COMSOL Multiphysics 3.5.  For time stepping, the Generalized-$\alpha$ method was used with the damping parameter $\rho_{inf}=1$.  A copy of the model can be obtained by contacting the corresponding author by email.

\begin{figure}[!t]
\begin{center}
\includegraphics[width=\textwidth]{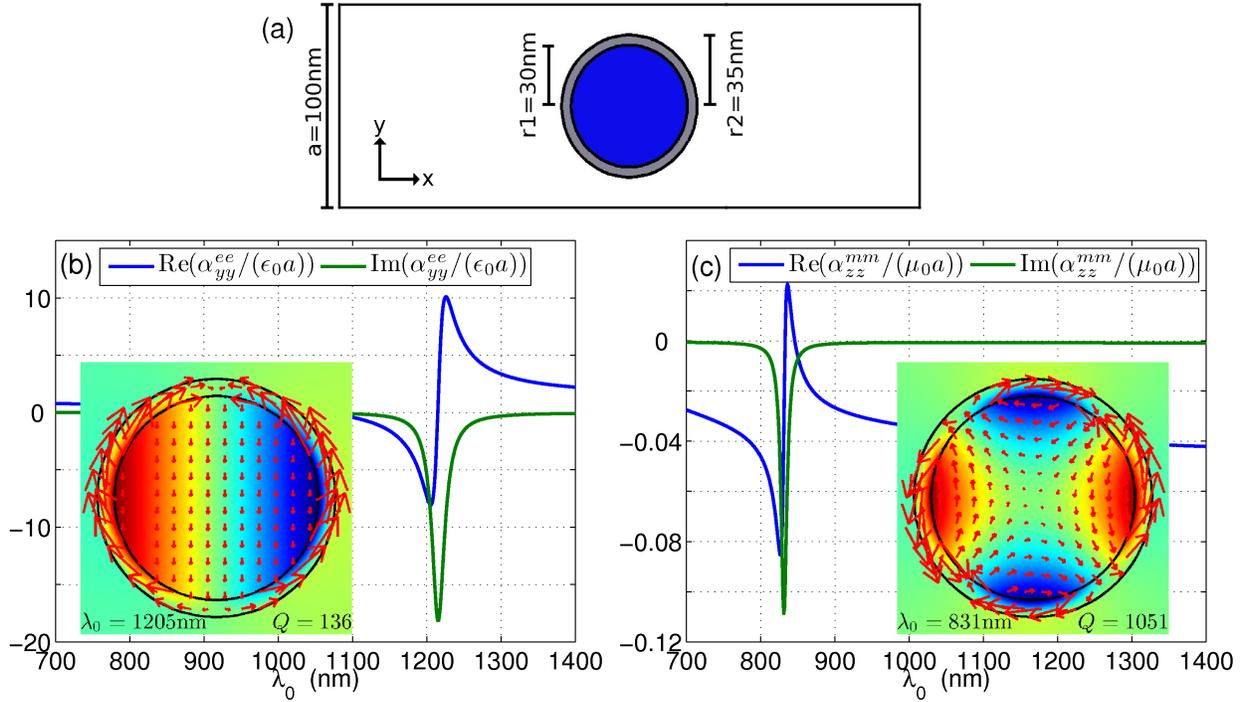}
\end{center}
\caption{(a) Diagram of the simulation domain for the one dimensional cylindrical spaser array with a core gain medium (blue) and outer Ag shell (gray).  A periodic boundary condition is imposed on the top and bottom boundaries, and a matched boundary condition (Sec.~\ref{Sec_2_3}) is imposed on the left and right boundaries.  Real and imaginary parts of the electric surface polarizability $\alpha_{yy}^{ee}$ (b) and magnetic surface polarizability $\alpha_{zz}^{mm}$ (c) are plotted, clearly indicating separate electric and magnetic resonnaces.  Inset are the field profiles for the two resonances and their corresponding wavelengths and Q factors.  Color indicates magnetic field $\mathrm{H}_z$, and arrows indicate the electric polarization $\textbf{P}=\textbf{D}-\textbf{E}$.}\label{Fig_2}
\end{figure}

The spaser is a one dimensional array of cylinders, each cylinder being infinite in extent in their axial direction.  Each cylinder has a core consisting of a four level gain medium with a radius of $r_1=30\mathrm{nm}$ and an outer shell composed of Ag with an outer radius of $r_2=35\mathrm{nm}$.  A diagram of the simulation domain is provided in Fig.~\ref{Fig_2}.  The artificial gain medium is characterized by the lifetimes $\tau_{10}=10^{-14}\mathrm{s}$, $\tau_{21}=10^{-11}\mathrm{s}$, $\tau_{32}=10^{-13}\mathrm{s}$ and $\tau_{30}=10^{-12}\mathrm{s}$.  The coupling constants in Eq.~(\ref{T_field3}) are $\sigma_a=10^{-4}\mathrm{C}^2/\mathrm{kg}$ and $\sigma_b=5\cdot10^{-6}\mathrm{C}^2/\mathrm{kg}$, and the linewidths of their corresponding transitions are $\Gamma_a=2\cdot10^{13}\mathrm{s}^{-1}$ and $\Gamma_b=1/\tau_{30}=10^{12}\mathrm{s}^{-1}$.  Finally, the initial population density parameter is $\mathrm{N}_{int}=5\cdot 10^{23}\mathrm{m}^{-3}$.  The population densities of the four level gain medium obeys the rate equations given in Eq.~(\ref{T_field4}), and the gain medium interacts with the electromagnetic field through the gain polarization which obeys Eq.~(\ref{T_field3}).  The Ag layer interacts with the electromagnetic field through the Drude polarization which evolves according to Eq.~(\ref{T_field2}).

Since the cylinder array is a single layer, it can be characterized as a metasurface~\cite{Kuester_03}.  As a metasurface, the electromagnetic response is given by the surface polarizability

\begin{equation}\label{alpha_value}
\begin{array}{c}
\hat{\alpha} = 
\left(\begin{array}{cc} 
\alpha_{yy}^{ee} & \alpha_{yz}^{em} \\
\alpha_{zy}^{me} & \alpha_{zz}^{mm}
\end{array}\right)
= -\displaystyle\frac{2\mathrm{i}}{\omega/c(1+S_{12}+S_{21}-\mathrm{det}(S))} \times \ \ \ \ \ \ \ \ \ \ \ \ \ \ \ \ 
\\ \\
\left(\!\!\!\begin{array}{cc}
\left[1+\mathrm{det}(S)-(S_{11}+S_{22})\right]\epsilon_0 & \left[(S_{12}-S_{21})-(S_{11}-S_{22})\right]/c \\[5pt]
\left[(S_{12}-S_{21})+(S_{11}-S_{22})\right]/c & \left[1+\mathrm{det}(S)+(S_{11}+S_{22})\right]\mu_0
\end{array}\!\!\!\right).
\end{array}
\end{equation}

\noindent Eq.~(\ref{alpha_value}) is adapted from ref~\cite{Fietz_10}, modified to be consistent with SI units and taking for granted that the metasurface is embedded in vacuum.  The surface polarizability $\hat{\alpha}$ is defined from the scattering matrix $S$.  The S matrix is defined from the amplitude of the electric field of the scattered waves and is adjusted so that the effective thickness of the characterized array is zero~\cite{Fietz_10}.  For a symmetric and reciprocal array, such as the cylindrical spaser array, the S matrix components $S_{11}=S_{22}$ are the reflection amplitude of a scattered wave and $S_{12}=S_{21}$ are the transmitted amplitude of the scattered wave.

The surface polarizability of the cylindrical array is plotted in Fig.~\ref{Fig_2}.  The reflection and transmission amplitudes used to calculate the surface polarizability were calculated from a frequency domain FEM simulation (COMSOL Multiphysics) where the Ag had a relative permittivity of $\epsilon_{Ag}=1-\omega_p^2/(\omega(\omega-\mathrm{i}\gamma))$ and the gain medium is simply a dielectric with permittivity $\epsilon_G=9$.  We see from the surface polarizabilities that there is an electric resonance near $\lambda_0=1220\mathrm{nm}$ and a magnetic resonance near $\lambda_0=830\mathrm{nm}$.  Fig.~\ref{Fig_2} also show fields profiles for each of these resonances calculated using a FEM eigenfrequency simulation.  Also shown are the wavelengths of the corresponding resonances $\lambda_r=2\pi c/\mathrm{Re}(\omega_r)$, and a resonance quality factor $\mathrm{Q}=2\pi\mathrm{Re}(\omega_r)/\mathrm{Im}(\omega_r)$, where $\omega_r$ is a complex eigenfrequency returned by the same FEM eigenfrequency simulation.

\begin{figure}[!t]
\begin{center}
\includegraphics[width=0.7\textwidth]{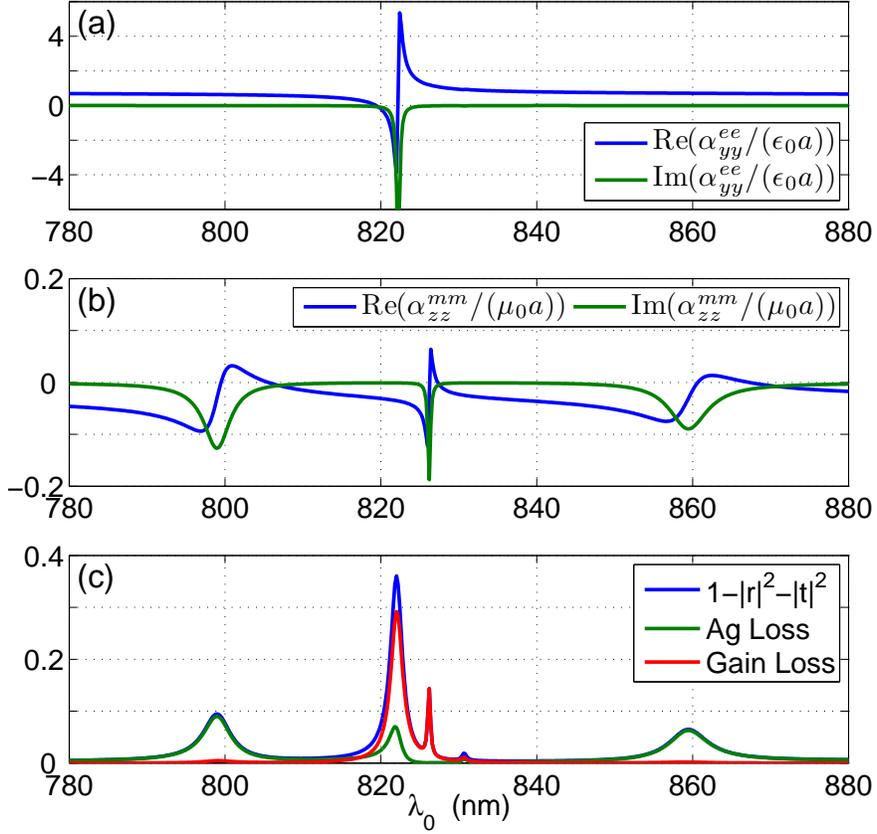}
\end{center}
\caption{(a) Electric surface polarizability $\alpha_{yy}^{ee}$ and (b) magnetic sufrace polarizability $\alpha_{zz}^{mm}$ for the cylindrical array with gain medium relative permittivity of $\epsilon_G=9-\sigma\mathrm{N}_{int}/(\omega^2-\omega_b^2-\mathrm{i}\Gamma_b\omega)$. (c) Total, absorption as well as absorption in Ag and absorption in the gain medium.  It is clear that the presence of the electronic transition in the gain medium strongly modifies the spectrum of the cylindrical array.}\label{Fig_3}
\end{figure}

We are interested in using both resonances to achieve lasing, one resonance for enhancing the pumping of the gain medium and the other resonance for enhancing the lasing transition.  Therefore we choose the energy levels of the artificial four level gain medium so that the $0 \rightarrow 3$ transition approximately matches the higher frequency magnetic resonance $\omega_b=2\pi c/830\mathrm{nm}$, and the $1 \rightarrow 2$ transition approximately matches the lower frequency electric resonance $\omega_a=2\pi c/1221\mathrm{nm}$.  The presence of a electronic transition will modify the spectrum of the cylindrical array for frequencies near that transition.  Fig.~\ref{Fig_3} plots the surface polarizability near the magnetic resonance at $\lambda_0=831\mathrm{nm}$ for the cylindrical array where the gain medium now has the relative permittivity $\epsilon_G=9-\sigma_b\mathrm{N}_{int}/(\omega^2-\omega_b^2-\mathrm{i}\Gamma_b\omega)$.  Fig.~\ref{Fig_3} also plots the total absorption of the cylindrical array, as well as separately plotting the absorption in the Ag and in the gain medium.  Like Fig.~\ref{Fig_2}, the data for these plots were calculated from a frequency domain FEM simulation.

We can see from Fig.~\ref{Fig_3} that the interaction of the electronic transition with the magnetic shape resonance shown in Fig.~\ref{Fig_2}(c) causes these resonances to hybridize.  As a result the response of the cylindrical array for frequencies near that transition is strongly modified.  Instead of a single magnetic resonance we see now see multiple resonances, both electric and magnetic.  Examining the absorption plotted in Fig.~\ref{Fig_3}(c) we see that the gain medium strongly absorbs at the magnetic resonance near $\lambda_0=826\mathrm{nm}$.  For our lasing simulations this will be the pump frequency.  There is no way to know exactly what the lasing frequency will be without first running the time domain lasing simulation, except to say that it will be approximately equal to the frequency of the $1 \rightarrow 2$ transition $\omega_a$.  A good initial guess is to set $\omega_1=\omega_a$, but after running the lasing simulation this can be adjusted to better match true lasing frequency.  In what follows, we have used $\omega_1=2\pi c/1219.3\mathrm{nm}$.

\begin{figure}[!t]
\begin{center}
\includegraphics[width=\textwidth]{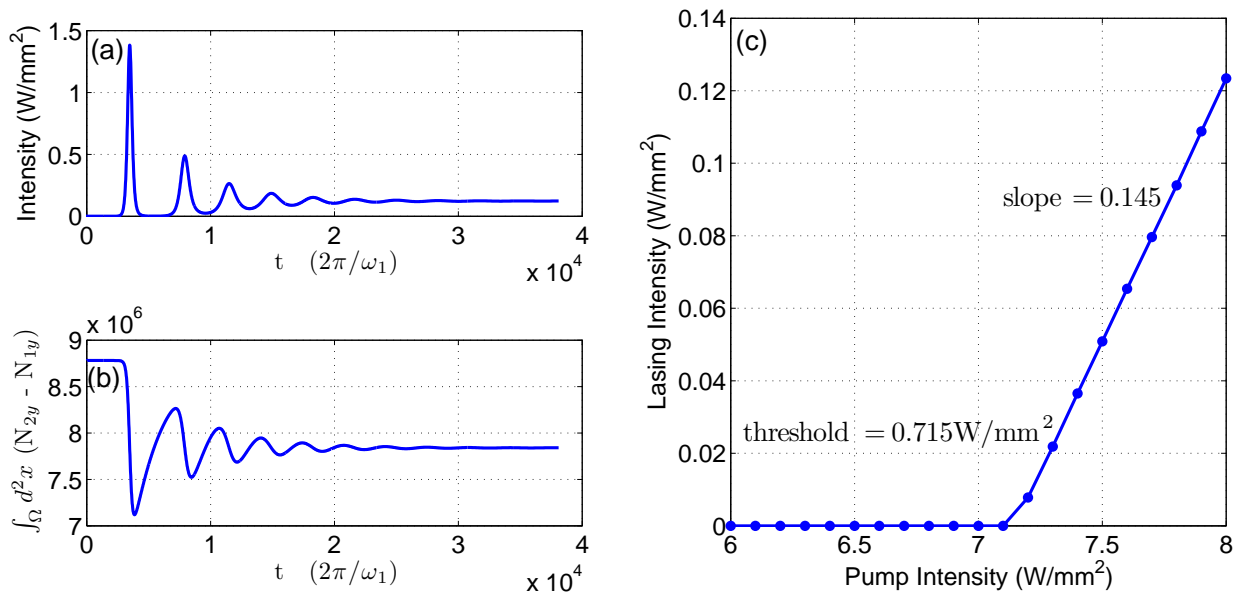}
\end{center}
\caption{(a) Lasing intensity defined as the power emitted outward from the array in either direction and (b) population inversion measured as $\int_{\Omega} d^2x (\mathrm{N}_{y2}-\mathrm{N}_{y1})$, where $\Omega$ is the domain of the simulation, and  as a function of time normalized to lasing periods.  The time domain simulation begins at $t=0$ with a steady state solution where the pump has been on for a very long time ($t\gg\tau_{21}$) and the system has population inversion without lasing due to the lack of spontaneous emmision.  (c)  Plot of steady state lasing intensity vs. pump intensity.  A linear fit indicates a pump threshold intensity of $7.15\mathrm{W/mm^2}$ and a slope of $0.145$.}\label{Fig_4}
\end{figure}

Fig.~\ref{Fig_4} shows data from a time domain simulation of the cylindrical spaser array using the parameters defined above.  The initial state of the simulation is prepared with a previous simulation where the system is pumped with the field $\textbf{A}_2$, with an intensity of $8\mathrm{W/mm^2}$, while the incident probe field is set to $\textbf{A}_1=0$.  The pump beam is turned on slowly with $\textbf{A}_2$ having the profile

\begin{equation}
\textbf{A}_2 = \mathrm{A}_{pu}\hat{\textbf{e}}_y \displaystyle\frac{1}{2}\left[1 + erf\left(\displaystyle\frac{t-5\tau_{pu}}{\sqrt{2}\tau_{pu}}\right)\right],
\end{equation}

\noindent where $\mathrm{A}_{pu}$ is the amplitude of the pump beam, $erf(x)=(2/\sqrt{\pi})\int_0^x e^{-t^2}dt$ is the error function, and $\tau_{pu}=1.0\cdot10^{-12}s$ is the pump rise time.  The pump beam excites oscillators in the gain medium to the third energy level, which decays to the second energy level at the rate of $1/\tau_{32}$.  After $t\gg \tau_{21}$, the system is in steady state population inversion, but cannot lase since our model does not allow for generation of light due to spontaneous emission.  The time is then reset, and the simulation shown in Fig.~\ref{Fig_4} begins in this steady state population inversion.  Shortly after t=0, a short probe pulse is emitted into the simulation domain.  This excites the polarization field $\textbf{P}_a^g$, which in turn begins the lasing process.  The intensity of the resulting lasing field plotted in Fig.~\ref{Fig_4}(a) spikes initially, but after about 30000 lasing periods it settles into steady state lasing.  Fig.~\ref{Fig_4} also plots the difference between the integral of the population densities $\mathrm{N}_2$ and $\mathrm{N}_1$ for the system beginning in population inversion.

The time step used for the simulations in Fig. 4 varies throughout the simulation.  When the pump is initially turned on the time step must be less then the pump rise time $\tau_{pu}$.  Once the pump is at a maximum the time step can be increased while the gain system approaches steady state.  When the time is reset and a probe pulse is introduced the time step must be made smaller than the width of the probe pulse, and must remain small to resolve the resulting oscillations of the interaction between the probe pulse and the resonators as well as the initial exponential growth of the lasing beam.  As the laser approaches steady state the time step can again be increased.  At all times the time step must be smaller than the inverse rate of change of any transient beams (pump, probe or lasing beams).  If $\omega_1$ is not close to the resulting lasing frequency the phase of $\textbf{A}_1$ will rapidly change and will require a correspondingly small time step.  Once the system begins lasing, the actual lasing frequency can be inferred from this oscillation in the phase of $\textbf{A}_1$, and $\omega_1$ can be changed in the middle of the simulation.  This causes the phase of $\textbf{A}_1$ to slowly change allowing for a larger time step.

There is a minimum pump intensity required for the light generated due to stimulated emission to overcome the internal losses in the cylindrical array.  Fig.~\ref{Fig_4}(c) plots the steady state lasing intensity vs. the pump intensity.  A linear fit to the lasing data points indicates a threshold pump intensity of $7.15\mathrm{W/mm^2}$.  This threshold intensity depends on a number of variables, including all of the parameters of the gain medium, as well as the cylinder plasmon resonances used to enhance both the pump and lasing transition (Figs.~\ref{Fig_2} and~\ref{Fig_3}).  These resonances in turn depend on the geometry and material parameters of the cylindrical array.

\begin{figure}[!t]
\begin{center}
\includegraphics[width=\textwidth]{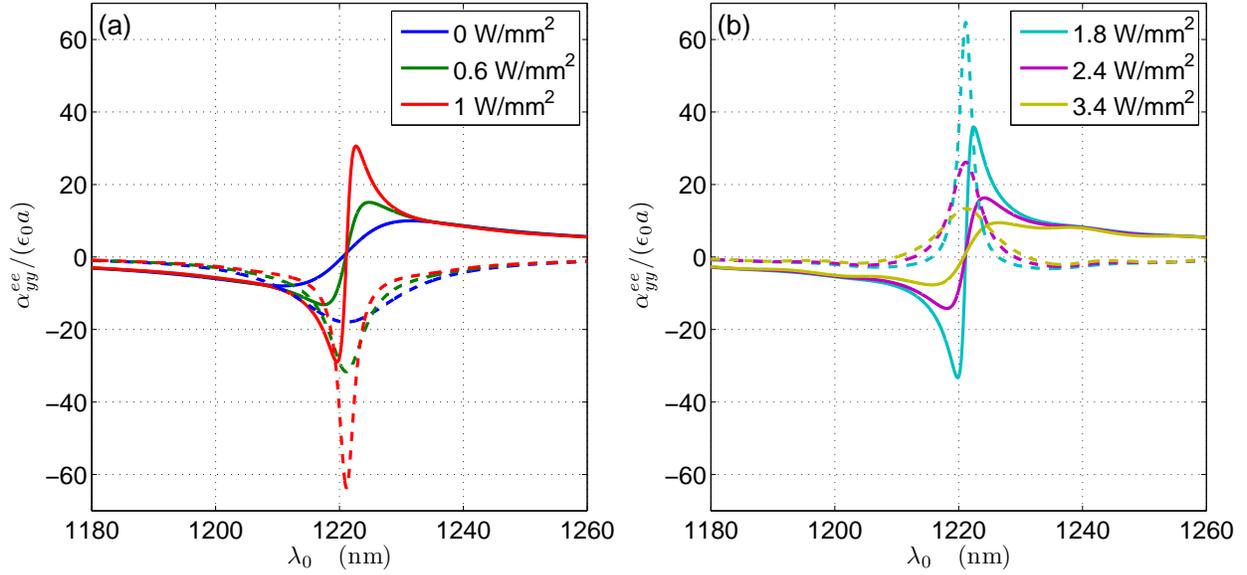}
\end{center}
\caption{Real (solid lines) and imaginary (dashed lines) parts of the electric surface polarizability $\alpha_{yy}^{ee}$ for the lasing electric resonance shown in Fig.~\ref{Fig_2} for various pump intensities (shown in legend).  In Fig.~\ref{Fig_5}(a) we see that at higher pump intensities the linewidth of the resonance narrows.  In Fig.~\ref{Fig_5}(b) we see that at even higher pump intensities the imaginary part of the surface polarizability flips (indicating gain) and the linewidth of the resonance begins to broaden.}\label{Fig_5}
\end{figure}

While there is a minimum threshold intensity for the array to exhibit lasing, we can observe interesting changes in the spectrum of the array at lower pump intensities.  We saw by comparing Figs.~\ref{Fig_2} and~\ref{Fig_3} that the spectrum of the cylindrical array was changed by the presence of the $0 \rightarrow 3$ transition.  As we pump the array at increasing intensities we observe a similar change in the spectrum due to the $1 \rightarrow 2$ transition.  Fig.~\ref{Fig_5} plots the surface polarizability (Eq.~(\ref{alpha_value})) of the electric resonance for different pump intensities.  These plots were created by pumping the cylindrical array with the field $\textbf{A}_2$ for a long period of time ($t\gg\tau_{21}$), and then injecting a Gaussian probe field $\textbf{A}_1$ with a much weaker intensity.  Applying a Fourier transform to the resulting time domain reflected and transmitted probe fields gives us the reflection and transmission amplitudes in the frequency domain, allowing us to calculate the surface polarizability according to Eq.~(\ref{alpha_value}).

From Fig.~\ref{Fig_5}, we see that for increasing values of the pump intensity, the lineshape of $\alpha_{yy}^{ee}$ resembles a Lorentzian

\begin{equation}
\alpha_{yy}^{ee} = \alpha_{inf} - \displaystyle\frac{\alpha_0}{\omega^2-\omega_{\alpha}^2-\mathrm{i}\gamma_{\alpha}\omega}.
\end{equation}

\noindent  We see in Fig.~\ref{Fig_5}(a) that as we increase the pump intensity, it is as if the positive valued scattering frequency $\gamma_{\alpha}$ is made smaller, narrowing the lineshape.  We see in Fig.~\ref{Fig_2}(b) that at even higher pump intensities, $\gamma_{\alpha}$ continues to shrink, passing through zero, and the imaginary part of $\alpha_{yy}^{ee}$ changes sign, indicating gain.  As the pump intensity continues to increase, $\gamma_{\alpha}$ continues to grow more negative and the lineshape begins to broaden.

Even though we have gain at the pump intensities in Fig.~\ref{Fig_5}(b), we still do not have lasing because the gain is not large enough to overcome radiative losses.

\section{Conclusion}

We have presented a finite element method simulation for a microphotonic lasing system.  We have shown how to achieve a massive speedup in the simulation by separating the various fields into fields that oscillate at the carrier frequencies $\omega_1$ or $\omega_2$, with slowly changing complex valued amplitudes.  A demonstration of this simulation was provided with a two dimensional model of a one dimensional cylindrical spaser array as an example.  The threshold pump intensity for this array was determined.  Finally, we have shown how the linewidth of the lasing transition changes for various pump intensities.

\section*{Acknowledgements}
Chris Fietz would like to acknowledge support from the IC Postdoctoral Research Fellowship Program.  Work at Ames Laboratory was supported by the Department of Energy (Basic Energy Science, Division of Materials Sciences and Engineering) under contract no. DE-ACD2-07CH11358.

\end{document}